\documentclass[aps,prl,twocolumn,showpacs,groupedaddress,amssymb]{revtex4}

\usepackage{graphicx} 

\begin{document}
\title{Direct experimental measurement of the speed-stress relation for dislocations in a plasma crystal}

\author{V. Nosenko}
\email{nosenko@mpe.mpg.de}
\author{G. E. Morfill}
\affiliation {Max-Planck-Institut f\"{u}r extraterrestrische Physik,
D-85741 Garching, Germany}

\author{P. Rosakis}
\affiliation {Department of Applied Mathematics, University of Crete, Heraklion 71409, Greece}

\date{\today}
\begin{abstract}
The speed-stress relation for gliding edge dislocations was experimentally measured for the first time. The experimental system used, a two-dimensional plasma crystal, allowed observation of individual dislocations at the ``atomistic'' level and in real time. At low applied stress dislocations moved subsonically, at higher stress their speed abruptly increased to $1.9$ times the speed of shear waves, then slowly grew with stress.  There is evidence that immediately after nucleation dislocations can move faster than pressure waves.
\end{abstract}
\pacs{
52.27.Lw, 
52.27.Gr, 
61.72.Ff,  
82.70.Dd 
} \maketitle

The viscoplastic behavior of crystalline solids is largely determined by the dynamics of dislocation motion under stress. A key hypothesis is the existence of a relation between dislocation speed and ambient shear stress.  Until recently, experimental measurements of this relation have reported dislocation speeds well below that of shear waves \cite{Johnston,Clifton:79}.  Theoretical models \cite{Rosakis:2001} and atomistic simulations \cite{Gumbsch:99} confirm that a speed-stress relation exists for uniformly moving dislocations; they also predict that at sufficiently high stress, dislocations move faster than shear waves, and possibly pressure waves.  This prediction has been controversial, since it contradicts  linear elastic models \cite{HL}. Direct experimental confirmation by tracking supersonically moving objects of atomistic size, would seem unlikely, unless an alternative model system were found. Plasma crystals \cite{Lin_I:96,Nosenko:07} are  a highly suitable candidate,  since  they can form a genuinely 2D,  millimeter-scale lattice comprised of micron-sized particles, with wave speeds of the order of millimeters per second. Motion of individual ``atoms''  (particles) is fully resolved  in real time, allowing direct observation of dislocation nucleation and dynamics  at the ``atomistic'' level.

A plasma crystal is a crystalline state of complex, or dusty plasma \cite{Thomas:94,Lin_I:94,Morfill:09}. Complex plasma is a suspension of fine solid particles in a weakly ionized gas \cite{Thomas:96,ShuklaBook,Fortov:05}.

Whereas the field of plasma crystals is rather mature, the study of dislocations in plasma crystals has only recently begun \cite{Lin_I:96,Knapek:07,Nosenko:07,Sheridan:08,Nosenko:09PRL,Durniak:10,Melzer:96,Feng:10}. Dislocation statistics was studied in Refs.~\cite{Knapek:07,Nosenko:09PRL}; the core energy was derived from the Arrhenius scaling of their concentration \cite{Nosenko:09PRL}.

Nucleation and dynamics of individual dislocations in a 2D plasma crystal were observed in \cite{Nosenko:07}. The shear stress  necessary for dislocation nucleation and motion was caused by naturally occurring slow differential rotation of the plasma crystal (presumably due to the ion drag force \cite{Konopka:00:Rotation}) hence could not be controlled on purpose. {\it Supersonic  dislocations} (faster than the shear wave speed $C_T$) were observed for the first time. The dislocation speed $v_{\rm disl}$ (measured from their Mach cone angles \cite{Nosenko:03}) was distributed in a narrow range of $(1.95\pm0.2)C_T$; the reason for this narrow distribution remained unclear.

In this Letter, we report the first direct experimental measurement of the glide speed of edge dislocations as a function of externally applied shear stress. Shear stress in a two-dimensional (2D) plasma crystal was applied in a controllable, homogeneous manner,  by a pair of counter-propagating laser beams. We observed subsonic and supersonic dislocations with a speed ``gap'' between them; weak dependence of dislocation speed on applied stress in the supersonic regime explains the narrow distribution of $v_{\rm disl}$ observed in \cite{Nosenko:07}. We find that supersonic dislocations play a central role in shear melting, and may be relevant in other systems, e.g. colloidal crystals \cite{Schall:06}, twinning \cite{Faran:2010}, as well as high-rate viscoplasticity \cite{HL}.

Laser manipulation is a very efficient method of creating relatively strong shear stress in a plasma crystal. The radiation pressure of a powerful laser beam in a suitable configuration is sufficient to shear-melt a plasma crystal and create a shear flow in it \cite{Nosenko:04PRL_visc}.

Shear melting of a plasma crystal naturally happens through increasing nucleation and proliferation of dislocations \cite{Thomas:96,Nosenko:09PRL}. For example, the $120^{\circ}$ zigzag features seen in \cite{Nosenko:04PRL_visc} in the particle velocity profiles at the onset of plastic deformation were most probably a signature of moving dislocations.

\begin{figure}
\centering
\includegraphics[width=80mm]{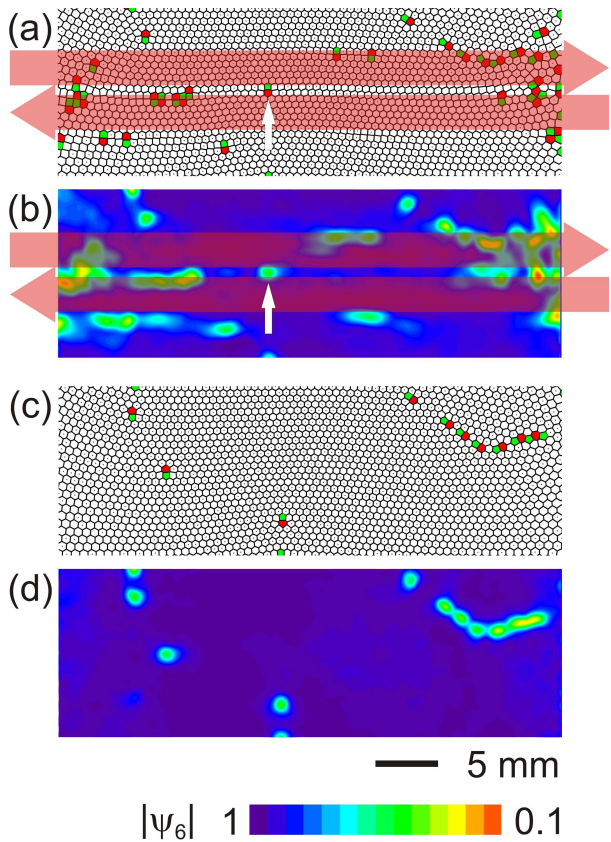}
\caption {\label {Maps} (a),(b) Shear stress is applied to a 2D plasma crystal by a pair of counter-propagating laser beams. Moving (to the right) dislocation is indicated by a small white arrow. (c),(d) The same crystal is shown before stress was applied. (a),(c) The dots are individual particles located inside their Voronoi cell, highlighted are lattice defects where the number of nearest neighbors is other than six. A dislocation is an isolated pair of five- and sevenfold lattice defects. (b),(d) To visualize the shear strain, we use the maps of the bond-orientational function $|\psi_6|$ \cite{Grier:94,Knapek:07,Nosenko:07}. The complete trajectory of this dislocation is shown in Fig.~\ref{displacement_t}(b).}
\end{figure}

To study dynamics of individual dislocations, with the aim of measuring the speed-stress relation, we implemented a new ``laser indenter'' configuration, schematically illustrated in Fig.~\ref{Maps}. New distinctive features of this configuration are the following: (1) Nearly homogeneous shear stress is created in a narrow gap with a width comparable to the interparticle spacing, between two wider stripes where the particles are pushed in opposite directions by the rastered laser beams and (2) one of the three axes of the triangular crystal lattice is aligned with the laser beams. This creates an experimental setting where dislocations nucleate and move in a clearly defined, essentially uniform and constant  shear-stress field.

Our experimental setup was a modified Gaseous Electronics Conference (GEC) rf cell as in \cite
{Couedel:10}, using similar experimental parameters. Argon plasma
was produced using a capacitively coupled rf discharge. The rf voltage had amplitude of $309$~V peak-to-peak at $13.56$~MHz. The self-bias voltage was $-125$~V.
To ensure that the system was not overdamped, a relatively low pressure of $5$~mTorr
was used. The neutral-gas damping rate is then accurately modeled
\cite {Liu:03} by the Epstein expression $\nu_E=\delta
N_gm_g\overline{v}_g(\rho_pr_p)^{-1}$, where $N_g$, $m_g$, and
$\overline{v}_g$ are the number density, mass, and mean thermal
speed of gas atoms and $\rho_p$, $r_p$ are the mass density and
radius of the particles, respectively. With leading coefficient
$\delta=1.26$ \cite {Liu:03}, this gave $\nu_E=0.77$~s$^{-1}$.

A monolayer of highly charged microspheres was levitated against
gravity in the sheath above the lower rf electrode. The particles
had a diameter of $9.19\pm0.09$~$\mu$m and a mass
$m=6.15 \times 10^{-13}$~kg. The monolayer, in the form of a triangular lattice, included about $6000$ particles and had a diameter of $\approx60$~mm.

The interparticle potential for particles arranged in a single plane, like ours, is well approximated \cite {Konopka:00:Yukawa} by the Yukawa potential: $U(r)=Q(4\pi\epsilon_0r)^{-1}{\rm exp}(-r/\lambda_D)$, where $Q$ is the particle charge and $\lambda_D$ is the screening length. The monolayer is characterized by a screening parameter $\kappa=\Delta/\lambda_D$, where $\Delta$ is the interparticle spacing. In our experiment, $\Delta=0.57$~mm. We used the spectral technique of \cite {Nunomura:02} to measure $\kappa=0.8\pm0.2$ and $Q=-15~000\pm1500e$.

The average sound speeds in the central part of our plasma crystal were measured
to be $C_L=34.2\pm 2.4$~mm/s and $C_T=6.1\pm 0.6$~mm/s, for
pressure and shear waves respectively. This gives a shear modulus $\mu=8.3\times 10^{-14}$~N/mm. Hexagonal lattice symmetry  ensures isotropy within the linear elastic regime.

The particles were imaged through the top window by a digital
camera. We recorded movies of $400$ frames at $60$ frames per
second. The $42.6\times42.6~\rm{mm}^2$ field of view included
$\approx 5100$ particles. The particle coordinates $x,y$ and
velocities $v_x,v_y$ were then calculated with subpixel resolution
for each particle in each frame.

In our ``laser indenter'' scheme, two oppositely directed laser beams are focused down to a fraction of the interparticle spacing and they are rapidly ($\simeq 300$~Hz) scanned to draw rectangular stripes on the suspension, as shown in Fig.~\ref{Maps}. The particles react to the averaged radiation pressure. A shear stress $\tau=n F/d$ is created in the $d=1.2\Delta$ wide gap between the laser-illuminated stripes, where $n$ is the number of lattice rows within each stripe and $F$ is the (averaged) radiation pressure force on each particle. The shear stress $\tau$ was controlled by varying the output laser power; they were proportional with the coefficient of $2.82\times 10^{-15}$ N/mm per $1$~W.

Assuming an initially quiescent linear elastic body, subjected to spatially constant opposite body forces in each stripe suddenly applied at $t=0$ and to (neutral-gas) drag body forces with the rate $\nu_E=0.77$~s$^{-1}$, we solved the equations of elastodynamics; this predicts that (i) after a rise time of roughly $h/C_T=0.6$~s ($h$ is the stripe width), the shear stress is nearly constant and maximum in the central unloaded gap between loaded stripes; (ii) neutral-gas drag  damps out  wave reflections. All our data are taken after this rise time.


When shear stress is suddenly applied, the plasma crystal first undergoes elastic deformation, then defect generation while in a solid state, and, if the stress is high enough, onset of plastic deformation and shear flow \cite{Nosenko:04PRL_visc}. Even before the visible onset of plastic deformation, dislocations nucleate and move in the lattice; their number rises significantly when laser power is increased.

Due to collinearity of the laser beams with a principal lattice axis, dislocations nucleate and move in the narrow gap between the laser-illuminated stripes. This substantially simplifies the analysis of their motion, which is nearly one-dimensional in the uniform and constant external shear-stress field.

Dislocations generally moved uniformly after nucleation; however, during a short initial period their speed was sometimes higher, see Figs.~\ref{displacement_t}(a),(b). Transitions from a high to a lower speed were usually well defined.

\begin{figure}
\centering
\includegraphics[width=80mm]{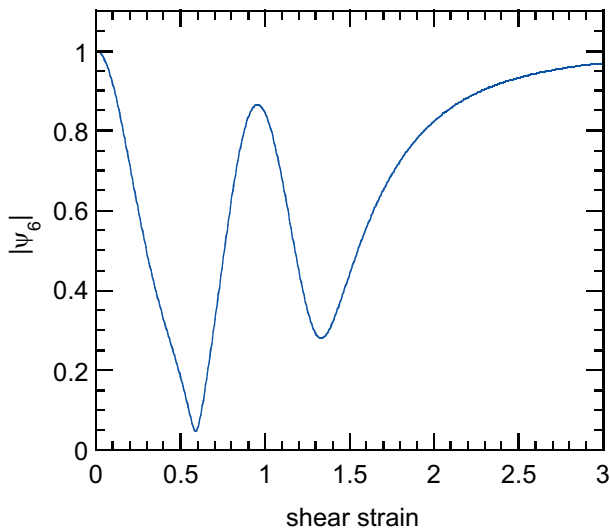}
\caption {\label {psi6_strain} Bond-orientational function $|\psi_6|$ versus simple shear strain $\gamma$ (along a principal axis of triangular lattice). Up to $\gamma=0.58$, the dependence is monotonic and can be used to calculate $\gamma$ from $|\psi_6|$ measurements. For small $\gamma$, this dependence reduces to $|\psi_6|=1-9\gamma^2$ \cite{Nosenko:07}.}
\end{figure}

\begin{figure}
\centering
\includegraphics[width=80mm]{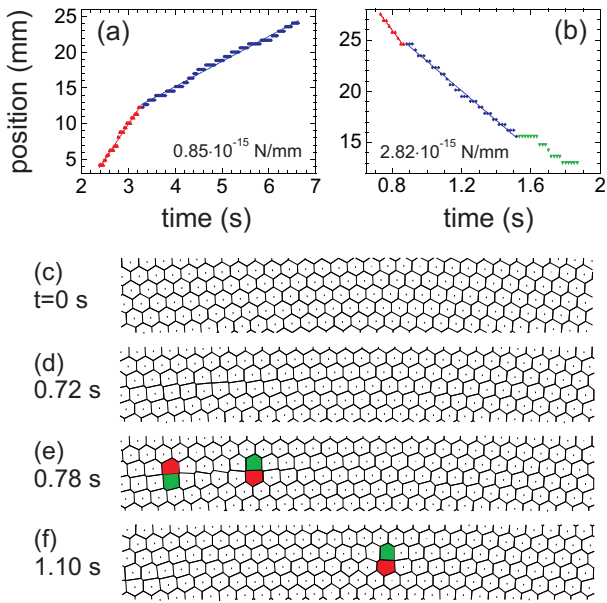}
\caption {\label {displacement_t} (a),(b) Positions of individual dislocations vs. time elapsed since shear stress $\tau$ was applied. The  value of $\tau$ is different   in each example as indicated. Voronoi diagrams are shown for the dislocation from (b) for four time instances: (c) before stress was applied, (d) just before nucleation, (e) during ``fast'' motion, and (f) during regular motion. Dislocation positions were calculated as the average positions of their constituent five- and sevenfold lattice defects.
}
\end{figure}

An explanation of this involves the behavior of shear strain. It can in principle be measured \cite{Nosenko:07} from the bond-orientational function $\psi_6$ \cite{Grier:94,Knapek:07}, shown in Figs.~\ref{Maps}(b),(d). For any lattice site, $\psi_{6}=\frac{1}{n}\sum^n_{j=1}{\rm exp}(6i\Theta_{j})$, where $\Theta_{j}$ are bond orientation angles for $n$ nearest neighbors. For small \textit{simple shear}, $|\psi_6|=1-9\gamma^2$, where $\gamma$ is the shear strain \cite{Nosenko:07}. For an arbitrarily large simple shear (along a principal lattice axis), we derived a relation between $|\psi_6|$ and $\gamma$ which reduces to the above result for small $\gamma$, see Fig.~\ref{psi6_strain}. Up to $\gamma=0.58$, the dependence is monotonic and can be used to calculate $\gamma$ from $|\psi_6|$ measurements.

We found that dislocations nucleated when a level of  $|\psi_6|\simeq0.35$, or $\gamma\simeq0.36$ was reached, with little regard to the laser power applied. The resulting behavior of $|\psi_6|$ is similar to that reported in \cite{Nosenko:07} (Fig.~3(a) therein) for ``spontaneously'' nucleating dislocations. This value of $\gamma\simeq0.36$  is quite a large strain, possibly associated with an unstable state. Lattice periodicity implies that shear stress  $\tau$ is a periodic function of $\gamma$, with period the lattice-invariant shear $\gamma_{inv}=1/\sqrt{3}\approx0.58$, such that a simple shear along a lattice axis maps a triangular lattice onto itself. Assuming e.g. a sinusoidal $\tau$-$\gamma$ relation, any strain between $\gamma_{inv}/4\approx0.14$ and $3\gamma_{inv}/4\approx0.43$ is unstable, e.g. $\gamma=\gamma_{inv}/2\approx0.29$ corresponds to a stacking fault, with atoms on adjacent planes right on top of each other.

\begin{figure}
\centering
\includegraphics[width=80mm]{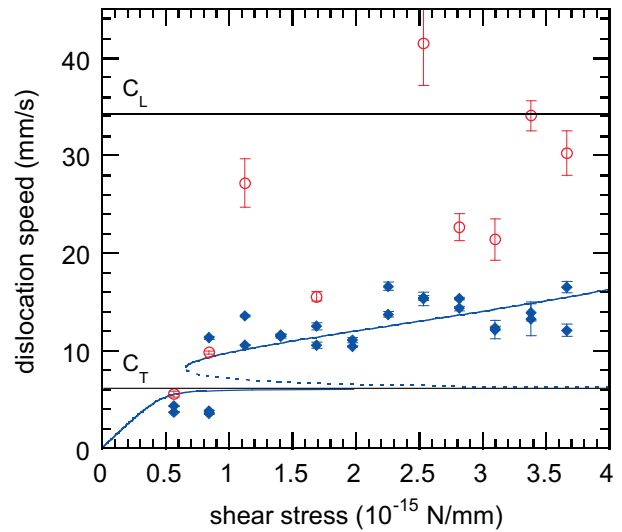}
\caption {\label {V_Stress} Dislocation speed as a function of applied shear stress. Open circles are for initial higher-speed motion (where it was present), solid diamonds for subsequent uniform motion. Data points for each setting of laser power were obtained, in most cases, in different experimental runs. The curve is a fit of the theoretical prediction of \cite{Rosakis:2001} to uniform-motion data (diamonds) only. Horizontal lines indicate the speeds of pressure and shear waves, $C_L$, $C_T$ respectively.
}
\end{figure}

This explains why some dislocations travel at a higher speed just after nucleation (open circles in Fig.~\ref{V_Stress}), then decelerate abruptly to a lower speed. A thin long region of high shear strain appears  along a glide plane as shown in Fig.~\ref{displacement_t}(d) (see also Fig.~1 of \cite{Nosenko:07}). From $|\psi_6|$ measurements, $\gamma$ in this region is in the range $0.3-0.4$, hence it probably is a stacking fault. Afterwards, a dislocation nucleates, moves into this region and annihilates it, then slows down. A model of Eshelby \cite{Eshelby:56} predicts that a dislocation moving into a stacking fault can propagate with arbitrary high speed above $C_L$. We sometimes observed this at high power settings. Thus there are two distinct types of dislocation motion: (i) motion into (and elimination of) a stacking fault [Fig.~\ref{displacement_t}(e) and open circles in Fig.~\ref{V_Stress}], and (ii) propagation into a stable lattice [Fig.~\ref{displacement_t}(f) and solid diamonds in Fig.~\ref{V_Stress}].  A calculation in the setting of \cite{Rosakis:2001} shows that at a given stress, a type (i) dislocation can move at speeds above $C_T$ but always faster than a type (ii). This explains the abrupt slowing down: when the fault is eliminated, a type (i) dislocation turns into a type (ii).   In the presence of thermal oscillations, the appearance  of unstable stacking faults, resulting in nucleation of fast dislocations that restore stability,  is more likely at higher stress.

Sometimes there is a departure from uniform motion at a later stage, which however has a simple explanation. For example, in Fig.~\ref{displacement_t}(b) the dislocation slowed down after (temporarily) merging with another dislocation at $t=1.5$~s. In some cases, dislocations slow down when entering a region where the crystal rows are not exactly straight. We used linear fits of dislocation position versus time at the stage of uniform motion to calculate the dislocation speed. The quality of these fits is good, except for the initial high-speed parts.

The main result of this Letter, the relation between dislocation speed and applied shear stress, is shown in Fig.~\ref{V_Stress}. Solid diamonds are for uniform motion of dislocations, open circles for occasional initial faster motion.


Our results are in qualitative agreement with theoretical models \cite{Rosakis:2001} and simulations \cite{Gumbsch:99}. The curve in Fig.~\ref{V_Stress} is a least-squares fit of the stress-speed relation predicted in  \cite{Rosakis:2001}  to our  uniform-motion data  (solid diamonds only). Between $C_T$ and $8.2$~mm/s (slightly below $\sqrt{2}C_T\approx 8.6$~mm/s) the speed-stress graph has negative slope, shown dashed in Fig.~\ref{V_Stress} and is thus unstable, which implies that supersonic dislocation speeds should be above $\simeq\sqrt2 C_T$, in agreement with our data. This  explains the speed gap between subsonic and supersonic dislocations  observed in Fig.~\ref{V_Stress}; see also \cite{Gumbsch:99}. Ref. \cite{Rosakis:2001} predicts that supersonic motion occurs above a critical stress (the turning point of the curve in Fig.~\ref{V_Stress}). The predicted value of this is $0.66\times 10^{-15}$~N/mm, compared to an observed value of $0.85\times 10^{-15}$~N/mm.
In the regime above $\simeq \sqrt{2}C_T$ a monotonically increasing relation is predicted between speed and shear stress. This is roughly the trend in our experiment. Weak dependence of $v_{\rm disl}$ on stress in this regime explains the narrow distribution of $v_{\rm disl}=(1.95\pm0.2)C_T$ observed in \cite{Nosenko:07}.  A  weak dependence of speed on stress also occurs in simulations  \cite{Gumbsch:99,Olmsted:05}. The fit yields a value for the theoretical shear strength of the crystal   of  $8.5\times 10^{-15}$~N/mm, roughly half the Frenkel estimate of  $\mu\gamma_{inv}/\pi\approx0.18\mu$.  It is also predicted that $v_{\rm disl}$ can exceed $C_L$ at stresses above $7.4\times 10^{-15}$~N/mm, well above the levels achieved in our experiments.

While there is a trend for dislocation speed to slowly increase with stress above $\sqrt{2}C_T$, data point scatter is appreciable. We discuss possible reasons. First, apart from neutral gas damping, the plasma crystal itself can behave viscoelastically. The ambient stress would then depend on time and it would matter when the measurement is taken. The latter is unlikely, however, since dislocations move more or less uniformly. There is the possibility though that the relation between dislocation speed and stress is itself history dependent in this particular system.

Second, stress can be inhomogeneous due to dislocations arrays (domain boundaries); these generate much longer-range stress fields than the localized self stress of  single dislocations. This affects moving dislocations.

Third, thermal atomic motion affects dislocation nucleation and propagation. Some energy from the laser forces eventually becomes ``thermal'' oscillations of the particles, in addition to their ``heating'' by fast ions streaming past particles from the plasma bulk to the rf electrode.

Fourth, nucleation or supersonic dislocation motion emits stress waves which might reflect from the boundary and alter the speed of dislocations they encounter. However, neutral gas damping dissipates these waves  before they reach the crystal edge, as predicted by our stress rise time calculation. Also, a simple estimate of the shear wave's damping length gives $C_T/\nu_E=7.9$~mm, much smaller than the crystal radius of $\approx30$~mm. Accordingly, the wings of observed Mach cones created by moving dislocations are much shorter than the crystal size.

There are two different regimes of dislocation motion above $C_T$. In the high-stress regime, $(2.3-3.7)\times 10^{-15}$~N/mm, one observes higher dislocation velocities on average, and a higher probability of initial fast motion, including occasional speeds above $C_L$ [this does not happen in the low-stress regime, $(1.1-2.0)\times 10^{-15}$~N/mm]. Another notable feature of the high-stress regime is  eventual shear melting (shear flow), absent from the low-stress regime. All data points in Fig.~\ref{V_Stress} are however taken from the first dislocations to emerge, before shear flow develops. Shear melting occurs above  a critical stress of $0.027\mu$, very close to values reported for various similar systems \cite{Zukoskiandothers}; it is caused by a proliferation of dislocations faster than $2C_T$. These observations point to supersonic dislocation motion as a key mechanism for shear melting.

We thank S. K. Zhdanov for valuable discussions.

\end{document}